\documentclass[Times,4pt,aps,pra,twocolumn,showpacs,amsmath,amssymb,floatfix,footinbib,superscriptaddress]{revtex4-2}
\usepackage{amsmath,natbib,graphicx,subfigure,amssymb,graphics,amsmath,mathrsfs,CJK,color,comment}
\usepackage{multirow,fancyhdr,color,bm,tabularx,psfrag,geometry,dcolumn,datetime,physics,booktabs}
\usepackage[colorlinks=true,linkcolor=blue,urlcolor=blue,citecolor=blue]{hyperref}
\usepackage[mathlines]{lineno}
\def\footnoterule{\kern -1mm \hrule width 5.8cm \kern 2.2mm}
\usepackage{tikz,xcolor,hyperref}
\definecolor{lime}{HTML}{A6CE39}
\DeclareRobustCommand{\orcidicon}{%
    \begin{tikzpicture}
    \draw[lime, fill=lime] (0,0)
    circle [radius=0.16]
    node[white] {{\fontfamily{qag}\selectfont \tiny ID}};\draw[white, fill=white] (-0.0625,0.095)
    circle [radius=0.007];
    \end{tikzpicture}
    \hspace{-2mm}}
\foreach \x in {A, ..., Z}
{\expandafter\xdef\csname orcid\x\endcsname{\noexpand\href{https://orcid.org/\csname orcidauthor\x\endcsname}{\noexpand\orcidicon}}}

\begin{document}
\title{Left handness in a four-level atomic system}
\author{Shuncai Zhao\orcidA{}}\email[ ]{zsczhao@126.com.}
\author{Zhengdong Liu}\email[Corresponding author: ]{lzdgroup@ncu.edu.cn.}

\affiliation{School of Materials Science and Engineering, Nanchang University Nanchang,Jiangxi Province, China,330031 ; 
Engineering Research Center for Nanotechnology,Nanchang University Nanchang Jiangxi Province, China,330047 ;  
Institute of Modern Physics, Nanchang University Nanchang Jiangxi Province,China,330031}

\begin{abstract}
A scheme is proposed for realizing simultaneous negative permittivity and negative permeability based on quantum coherence in
a four-level dense atomic system here.Under some parametric conditions the system shows that simultaneous negative permittivity
and negative permeability(i.e.Left handness) can be achieved in a wider frequency band because of quantum coherence.And the novelty
properties of gain and dispersion near the resonance frequency may have some potential applications.
\begin{description}
\item[PACs]{42.50.Gy}
\item[Keywords]{Quantum coherence; Negative refraction index; Left handness; Atomic system.}
\end{description}
\end{abstract}
\maketitle

\section{Introduction}
The propagation of electromagnetic waves in matter is characterized by the frequency-dependent relative dielectric permittivity
$\varepsilon_{r}$ and magnetic permeability $ \mu_{r}$.And their product defines the index of refraction:$\varepsilon_{r}\cdot\mu_{r}=n^{2}$. The left-handed
material(LHM) has a negative refractive index when the permittivity and permeability are negative simultaneously[1].It is a new kind of
materials that offers a possibility of molding the flow of light inside media,and it has attracted considerable attention[2-13]
because of its surprising and counterintuitive electromagnetic and optical properties.In LHM,the wave vector is opposite to the
direction of energy propagation.In other words, the Poynting vector and the wave vector of electromagnetic wave would be antiparallel,
i.e., the vector $\vec{k}$, electric field $\vec{E}$ and magnetic field $ \vec{H}$ form a left-handed system; thus Veselago referred to
such materials as``left-handed" media[1],and correspondingly,the conventional media in which $\vec{k}$,$\vec{E}$ and $ \vec{H}$ form
a right-handed system may be termed the ``right-handed" materials. It can be readily verified that the left-handed media exhibits a number
of peculiar electromagnetic and optical properties,such as the reversals of both Doppler shift and Cherenkov radiation[1],anomalous
refraction, negative Goos-H\"{A}anchen shift[5],reversed circular Brag phenomenon[12],amplification of evanescent waves[7], enhanced
the quantum interference[33], modified spontaneous emission rates[34] as well as unusual photon tunneling effect[6], subwavelength 
focusing[7,13], and so on[35]. Although naturally occurring materials with both negative permittivity and permeability simultaneously 
are not available, the left-handed materials have been realized by several approaches, including
artificial composite metamaterials[14-17], transmission line simulation[18], photonic crystal structures[11,19-20],and chiral
material[21-23]as well as photonic resonant materials[24-27]. The former four methods,based on the classical electromagnetic theory,
require delicate manufacturing of spatially periodic structure.The last method is a quantum optical approach where the physical
mechanism is the quantum interference and coherence that arises from the transition process in a multilevel atomic system.It was first
simultaneously proposed by Oktel et al. in a three-level medium[24], which requires rigorous level condition.For the purposes
of enhancing the freedom of choice of levels and making the scheme much more applicable to realistic system, Thommen et al.presented a 
proposal based on a coherent cross-coupling between electric and magnetic dipole transitions in a four-level scheme[27].

In this paper,we develop the mechanism and theoretically put forward a four-level dense atomic vapor scheme based on quantum coherence
effect to realize left-handness.In our scheme,the contribution of Lorentz-Lorenz local field of the four-level dense atomic vapor
should be considered.The system shows simultaneously negative permittivity and negative permeability under some appropriate
conditions(concerning the atomic electric and magnetic transition dipole moments, spontaneous decay rates and dephasing rates of
atomic levels,intensities of the external coupling and signal fields as well as the atomic concentration of the coherent atomic vapor)in
a wider frequency band than that of Ref.[27],and shows gain, novelty dispersion properties at certain probe frequency bands.
\section{Four-level atomic system's electric and magnetic polarization}

Consider a four-level configuration atomic ensemble interacting with three optical fields,i.e. the coupling beam,probe light and signal
field, and the frequency detunings of these three optical fields are $\Delta_{c}$,$\Delta_{p}$ and $\Delta_{s}$,respectively.The
configuration of such a four-level system is depicted in Fig.1.The Rabi frequencies of these optical fields are denoted by
$\Omega_{c}$,$\Omega_{p}$ and $\Omega_{s}$, respectively.The two levels,$|2>$ and $|3>$ have the same parity and
$\mu_{23}=<2|\vec{{\mu_{23}}}|3>$,where $\vec{{\mu_{23}}}$ is the magnetic dipole operator, and the levels $|2>$ and $|4>$ have an odd
parity with $d_{24}=<2|\vec{d_{24}}|4>$,where $\vec{d_{24}}$ is the electric dipole operator.$\gamma_{ij}$ are the population decay
rates from level i to j and $\Gamma_{ij}$ are the dephasing rates of the transition between levels i and j.

\begin{figure}[ph]
\centering
\includegraphics[width=2.0in]{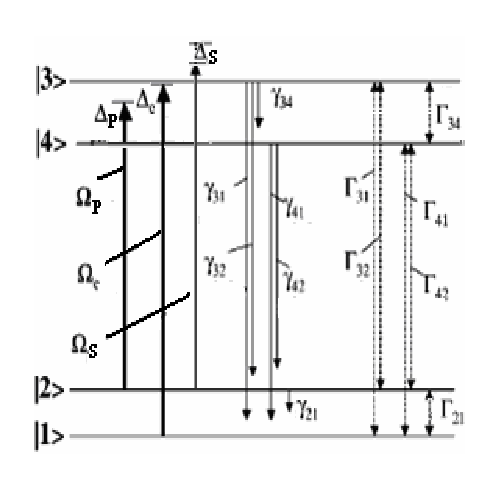}
\caption{The schematic diagram of a four-level dense atomic system. The level pairs$|2\rangle-|4\rangle,|2\rangle-|3\rangle$ are coupled
to the electric and magnetic fields of the probe light,respectively. The electric-dipole transitions$|1\rangle-|3\rangle$ and $ |2\rangle-|4\rangle$ are driven by the strong coupling andsignal laser beams,respectively.}\label{fig1}
\end{figure}

Using the density-matrix approach, the time-evolution of the system is described as
\begin{equation}
\frac{d\rho}{dt}=-\frac{i}{\hbar}[H,\rho]+\Lambda\rho ,\label{eq1}
\end{equation}
Where $\Lambda\rho$ represents the irreversible decay part in the system which is a phenomenological added decay term that corresponds
to the incoherent processes.The density matrix equations are given as,
\begin{align}
\dot{\rho_{11}}&=\gamma_{14}\rho_{44}+\gamma_{13}\rho_{33}+\gamma_{12}\rho_{22}-i\Omega_{c}(\rho_{31}-\rho_{13})\\
\dot{\rho_{22}}&=\gamma_{24}\rho_{44}+\gamma_{23}\rho_{33}-\gamma_{12}\rho_{22}+\gamma_{21}\rho_{11}-i\Omega_{s}(\rho_{32}-\rho_{23})\\
\dot{\rho_{33}}&=-\gamma_{34}\rho_{33}-\gamma_{13}\rho_{33}-\gamma_{23}\rho_{33}-i\Omega_{c}(\rho_{31}-\rho_{13})-i\Omega_{s}(\rho_{32}-\rho_{23})\\
\dot{\rho_{44}}&=-\gamma_{24}\rho_{44}-\gamma_{14}\rho_{44}+\gamma_{34}\rho_{33}+i\Omega_{p}(\rho_{24}-\rho_{42})\\
\dot{\rho_{12}}&=-(\Gamma_{12}+i\Delta_{12})-i\Omega_{c}\rho_{23}-i\Omega_{s}\rho_{31}-i\Omega_{p}\rho_{41}\\
\dot{\rho_{13}}&=-(\Gamma_{13}+i\Delta_{13})\rho_{13}-i\Omega_{c}(\rho_{11}-\rho_{33})-i\Omega_{s}\rho_{12}\\
\dot{\rho_{14}}&=-(\Gamma_{14}+i\Delta_{14})-i\Omega_{p}\rho_{12}-i\Omega_{c}\rho_{43}\\
\dot{\rho_{23}}&=-(\Gamma_{23}+i\Delta_{23})\rho_{23}-i\Omega_{s}(\rho_{22}-\rho_{33})-i\Omega_{p}\rho_{43}-i\Omega_{c}\rho_{21}\\
\dot{\rho_{24}}&=-(\Gamma_{24}+i\Delta_{24})\rho_{24}-i\Omega_{p}(\rho_{22}-\rho_{44})-i\Omega_{p}\rho_{34}\\
\dot{\rho_{34}}&=-(\Gamma_{34}+i\Delta_{34})\rho_{34}-i\Omega_{p}\rho_{32}-i\Omega_{s}\rho_{42}-i\Omega_{c}\rho_{41}\
\end{align}

Where$\Delta_{12}=\Delta_{c}-\Delta_{s},\Delta_{14}=\Delta_{c}-\Delta_{p}-\Delta_{s},\Delta_{34}=\Delta_{p}-\Delta_{s}$.
In the following,we will discuss the electric and magnetic responses of the medium to the probe field.When discussing how the detailed
properties of the atomic transitions between the levels are related to the electric and magnetic susceptibilities, one must make a
distinction between macroscopic fields and the microscopic local fields acting upon the atoms in the vapor. In a dilute vapor,there is
little difference between the macroscopic fields and the local fields that act on any atoms(molecules or group of
molecules)[28].But in dense media with closely packed atoms(molecules), the polarization of neighboring atoms(molecules)
gives rise to an internal field at any given atom in addition to the average macroscopic field, so that the total fields at the atom are
different from the macroscopic fields[28].In order to achieve the negative permittivity and permeability, here the chosen vapor with
atomic concentration$N=5\times10^{24}m^{-3}$ should be dense, so that one should consider the local field effect, which results from the
dipole-dipole interaction between neighboring atoms.In what follows we first obtain the atomic electric and magnetic polarizabilities,
and then consider the local field correction to the electric and magnetic susceptibilities(and hence to the permittivity and
permeability)of the coherent vapor medium. With the formula of the atomic electric polarizations
$\gamma_{e}=2d_{24}\rho_{24}/\epsilon_{0}E_{p}$,where$E_{p}=\hbar\Omega_{p}/d_{24}$
one can arrive at
\begin{eqnarray}
\gamma_{e}=\frac{2d_{24}^2\rho_{42}}{\epsilon_{0}\hbar\Omega_{p}}\
\end{eqnarray}
 In the similar fashion, by using the formulae of the atomic magnetic polarizations$\gamma_{m}=2\mu_{0}\mu_{12}\rho_{21}/B_{p}$ [28],and
the relation of between the microscopic local electric and magnetic fields$E_{p}/B_{p}=c$ we can obtain the explicit expression for the
atomic magnetic polarizability.Where $\mu_{0}$is the permeability of vacuum,c is the speed of light in vacuum.Then,we have obtained the
microscopic physical quantities $\gamma_{e}$and$\gamma_{m}$ . In order to achieve a significant magnetic response,it should be noted
that the transition frequency between levels$|2>-|3>$,and $|2>-|4>$ should be approximately equal to the frequency of the probe
light.Thus,the coherence $\rho_{23}$ drives a magnetic dipole, while the coherence $ \rho_{24}$ drives an electric dipole.This,therefore,
means that levels $|3>$and$|4>$ are very close to each other is required.However,what we are interested in is the macroscopic
physical quantities such as the electric and magnetic susceptibilities which are the electric permittivity and magnetic
permeability.The electric and magnetic Clausius-Mossotti relations can reveal the connection between the macroscopic and microscopic
quantities. According to the Clausius-Mossotti relation [28], one can obtain the electric susceptibility of the atomic vapor medium
\begin{eqnarray}
\chi_{e}=N\gamma_{e}\cdot{{{{(1-\frac{N\gamma_{e}}{3})}}}}^{-1}
\end{eqnarray}
The relative electric permittivity of the atomic medium reads $\varepsilon_{r}=1+\chi_{e}$.In the meanwhile,the magnetic
Clausius-Mossotti[29]
\begin{eqnarray}
\gamma_{m}=\frac{1}{N}(\frac{\mu_{r}-1}{\frac{2}{3}+\frac{\mu_{r}}{3}})
\end{eqnarray}
shows the connection between the macroscopic magnetic permeability $\mu_{r}$ and the microscopic magnetic polarizations $\gamma_{m}$. It
follows that the relative magnetic permeability of the atomic vapor medium is
\begin{eqnarray}
\mu_{r}=\frac{1+\frac{2}{3}N\gamma_{m}}{1-\frac{1}{3}N\gamma_{m}}
\end{eqnarray}
In the above,we obtained the expressions for the electric permittivity and magnetic permeability of the coherent atomic vapor
medium.In the section that follows,we will demonstrate that under the appropriate conditions the permittivity and permeability of the
four-level coherent atomic vapor medium can be simultaneously negative at certain probe frequency ranges.

\section{Results and discussion}

\begin{figure} \centering
\includegraphics[width=4.0in]{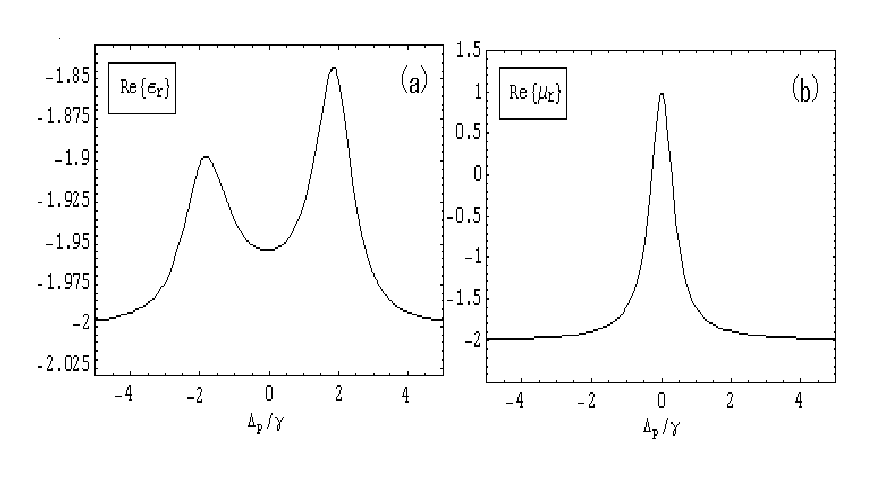}
\caption{Frequency dependence of the real parts of relative
dielectric permittivity and magnetic permeability on the probe field detuning$\frac{\Delta p}{\gamma}$ The typical parameters are chosen as:$\Omega_{s}=3.80\gamma$,$\Omega_{c}=1.80\gamma$,$\Delta_{s}=0.0001\gamma$,$\Gamma_{34}=0.01\gamma$,$\Gamma_{42}=0.006\gamma$, $\Gamma_{32}=0.005\gamma$,$\gamma_{34}=0.0076\gamma$,$\Gamma_{21}=\Gamma_{34}=\Gamma_{41}=0.0001\gamma$,$\gamma_{42}=\gamma_{32}=\gamma_{21}=\gamma_{13}=\gamma_{41}=0.00018\gamma$,$\gamma=10^{10}$}.\label{fig2}
\end{figure}

We consider the coupling laser beam is assumed to be on resonance, i.e.,the frequency detunings $\Delta_{c}=0$,and the atomic vapor
medium with $N=5\times10^{24}m^{-3}$to examine the case of dense media where the Lorentz-Lorenz local field corrections play a
significant role.As pointed out in the following numerical example, the Rabi frequency of the applied coherent optical field is
$(10^{10}-10^{11}/s)$,which is larger than the dephasing rate. Thus, the effects of dephasing, and collisional broadening can be
neglected,and the coherence effect can be maintained even increased in the atomic vapor under consideration.At room temperature the
atomic level shift associated with the Doppler effect is $(10^{7}-10^{8}/s)$.As shown in the following example(see Figs.2,3
and 4),the negative-index frequency detuning can be in the range of $(10^{10}/s)$, which is larger than the effect of the Doppler
broadening.This means that the Doppler broadening even at room temperature cannot affect the present numerical results and that our
calculation is self-consistent.Therefore,we had neglected the Doppler shift in the equations of motion of the density matrix.With
the steady solution of the matrix equations(2)we can obtain the coherent terms $\rho_{23}$ and $\rho_{24}$. And with the expressions 
for the atomic electric and magnetic polarizabilities(4)-(7),we will
present a numerical example to show that the strong electric and magnetic responses can truly arise in the four-level coherent vapor
medium under certain appropriate conditions. The strong electric and magnetic responses can even lead to simultaneously negative
permittivity and permeability at a certain wide frequency bands of the probe light.The real and imaginary parts of the relative
permittivity and permeability are plotted in Figs.2 and Figs.3, respectively.

\begin{figure}
\centering
\includegraphics[width=3.0in]{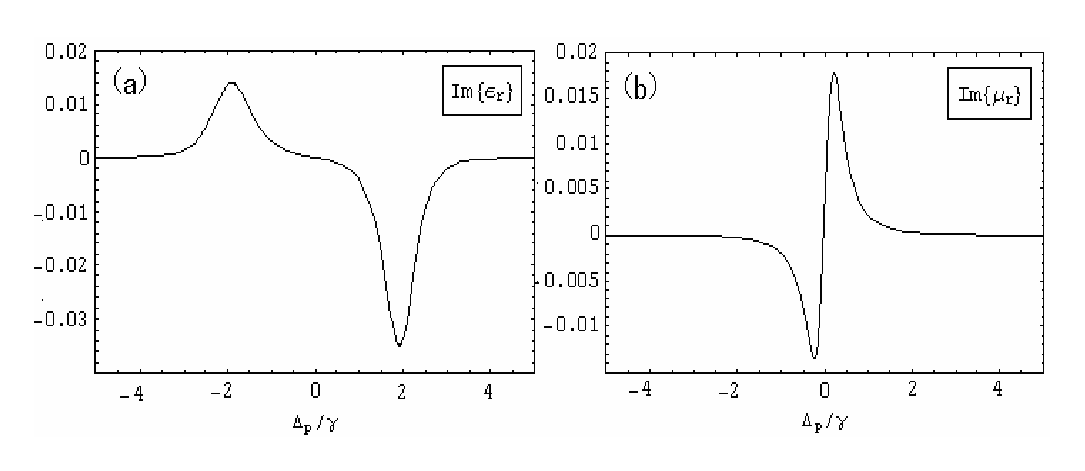}
\caption{Frequency dependence of the imaginary parts of relative dielectric permittivity and magnetic permeability on the probe field detuning$\frac{\Delta p}{\gamma}$,The typical parameters are chosen
the same as in Fig.2.}\label{fig3}
\end{figure}

It can be seen from Fig.2(a)that the relative dielectric permittivity has a negative real part in the probe frequency
detuning ranges[$-5\gamma,5\gamma$],and Fig.2(b)exhibits that the real part of relative magnetic permeability is negative in the
ranges[$-5\gamma,-0.5\gamma$]and [$0.5\gamma,5\gamma$]which are wider than that of the Ref.[26],and a positive peak approximately in
the range[$-0.5\gamma,0.5\gamma$].So,the four-level coherent atomic vapor can exhibit simultaneously negative permittivity and
permeability since the real part of the relative permittivity in these frequency detuning ranges is negative.

Fig.3 gives the relation of the imaginary parts of relative dielectric permittivity and magnetic permeability on the probe field
detuning.In Fig.3(a)the probe light exhibits electromagnetically induced transparency, a weak absorption peak at the values
$-1.92\gamma$and a stronger gain weak at the values $ 1.92\gamma$ of the probe field detuning .The imaginary part of the relative
permeability shows a negative peak at the probe field detuning $-0.261\gamma$,a positive peak at $0.192\gamma$ ,zero at other
ranges in Fig.3(b).

\begin{figure}
  \centering
  \includegraphics[width=3.0in]{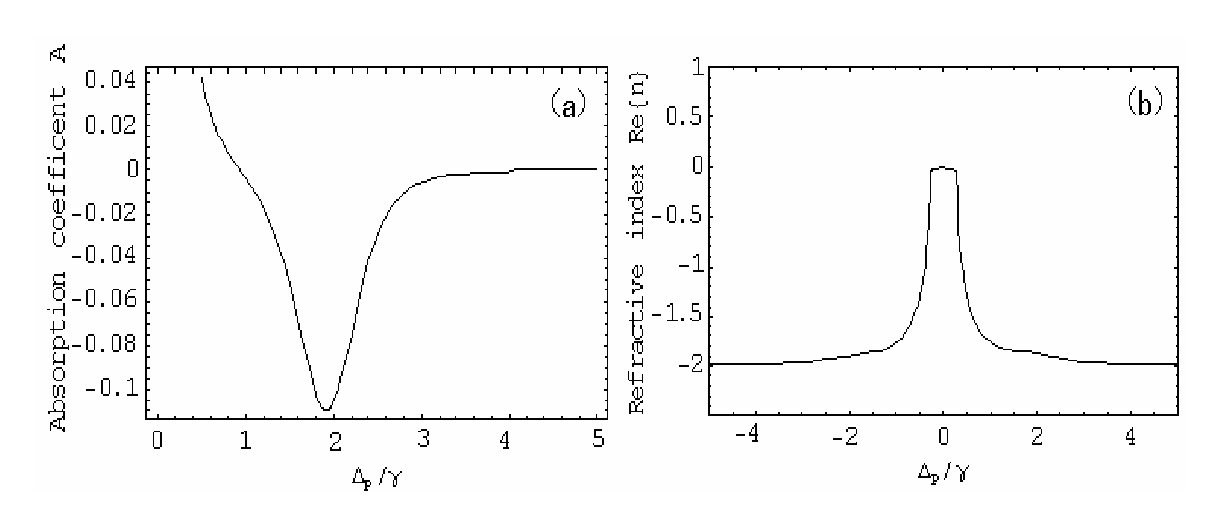}
 \caption{(a)Dependence of the absorption coefficient A on the probe field detuning $\frac{\Delta p}{\gamma}$,(b)Dependence of the
refractive index Re$\{n\}$on the probe field detuning $\frac{\Delta p}{\gamma}$ The typical parameters are chosen the same as in
Fig.2.}\label{fig4}
\end{figure}

The absorption property of the atomic vapor media is depicted in Fig.4(a),the absorption coefficient of LHM is expressed as
$A=2\pi{Im[n_{r}]}=2\pi{Im[-\sqrt{\varepsilon_{r}\mu_{r}}}]$, which is different to the case of an ordinary medium
$A=2\pi{Im[-\sqrt{\varepsilon_{r}}}]$.In Fig.4(a)the probe field detuning ranges are [0,5$\gamma$].In the probe frequency detuning
range [0.915$\gamma$,5$\gamma$],the absorption coefficient is negative and its negative peak appears in the range
of[1.8$\gamma$,2.0$\gamma$].In other words,the LHM can be no absorption but gain in some of the probe frequency detuning
ranges.Maybe our scheme is a potential application in high resolution imaging and beam refocusing,because the main applied
limitation of the left-handed materials is the large amount of dissipation and absorption[29].Particularly,the resolution of
perfect lens[7]is obviously decreased because of the absorption.Fig.4(b)shows the dispersion properties of the vapor
media.The dispersion approximately vanishes in the range[-0.5$\gamma$,0.5$\gamma$] of the probe beam.But it exhibits
different dispersive properties at two sides of it, namely, normal dispersion in the range[-5$\gamma$,-0.5$\gamma$]and anomalous
dispersion[0.5$\gamma$,5$\gamma$]of the probe frequency detuning.According to the group velocity definition
$v_{g}=\frac{c}{Re[(n+\omega\frac{dn}{d\omega})]}$,the superluminal propagation will occur in [0.5$\gamma$,5$\gamma$], the subluminal
propagation in [-5$\gamma$,-0.5$\gamma$].Therefore, maybe we can manipulate the probe beam to be from superluminal to subluminal or
vice versa in this left-handed material.

\section{Conclusion}

In this paper, the possibility of realizing the negative refractive index in a multilevel atomic system was studied by using the quantum
coherence effect.The simultaneously negative permittivity and negative permeability were achieved at a wider probe frequency bands
in such a coherent atomic vapor system under appropriate conditions. it can be flexible to manipulate of the negative refractive index by
using the external fields (e.g.,signal and controlling fields). And the gain properties of the LHM may be a scheme to solve the main
applied limitation of the left-handed materials because of the dissipation and absorption, especially in perfect lens[7].It May has
potential applications in improvements of the perfect lens resolution[7,31], beam focusing[32],and so on.

\section*{Acknowledgments}

The work is supported by the National Natural Science Foundation of China ( Grant No.60768001 and No.10464002 ).

\bibliographystyle{apsrev4-2}
\end{document}